\documentclass[%
reprint,
superscriptaddress,
 amsmath,amssymb,
 aps,
 longbibliography
]{revtex4-1}

\usepackage{graphicx}
\usepackage{dcolumn}
\usepackage{bm}
\usepackage{color}

\begin{document}

\preprint{APS/123-QED}

\title{Sampling networks by nodal attributes}

\author{Yohsuke Murase}
\email{yohsuke.murase@gmail.com}
\affiliation{RIKEN Center for Computational Science, 7-1-26, Minatojima-minami-machi, Chuo-ku, Kobe, Hyogo, 650-0047, Japan}
\author{Hang-Hyun Jo}
\affiliation{Asia Pacific Center for Theoretical Physics, Pohang 37673, Republic of Korea}
\affiliation{Department of Physics, Pohang University of Science and Technology, Pohang 37673, Republic of Korea}
\affiliation{Department of Computer Science, Aalto University, Espoo FI-00076, Finland}
\author{J\'anos T\"or\"ok}
\affiliation{Department of Theoretical Physics, Budapest University of Technology and Economics, H-1111 Budapest, Hungary}
\affiliation{Department of Network and Data Science, Central European University, N\'ador u.~9, H-1051 Budapest, Hungary}
\affiliation{MTA-BME Morphodynamics Research Group, Budapest University of Technology and Economics, H-1111 Budapest, Hungary}
\author{J\'anos Kert\'esz}
\email{KerteszJ@ceu.edu}
\affiliation{Department of Network and Data Science, Central European University, N\'ador u.~9, H-1051 Budapest, Hungary}
\affiliation{Department of Computer Science, Aalto University, Espoo FI-00076, Finland}
\author{Kimmo Kaski}
\affiliation{Department of Computer Science, Aalto University, Espoo FI-00076, Finland}
\affiliation{The Alan Turing Institute, British Library, 96 Euston Road, London NW1 2DB, UK}


\date{\today}

\begin{abstract}
In a social network individuals or nodes connect to other nodes by choosing one of the channels of communication at a time to re-establish the existing social links. Since available data sets are usually restricted to a limited number of channels or layers, these autonomous decision making processes by the nodes constitute the sampling of a multiplex network leading to just one (though very important) example of sampling bias caused by the behavior of the nodes. We develop a general  setting to get insight and understand  the class of network sampling models, where the probability of sampling a link in the original network depends on the attributes $h$ of its adjacent nodes. Assuming that the nodal attributes are independently drawn from an arbitrary distribution $\rho(h)$ and that the sampling probability $r(h_i , h_j)$ for a link $ij$ of nodal attributes $h_i$ and $h_j$ is also arbitrary, we derive exact analytic expressions of the sampled network for such network characteristics as the degree distribution, degree correlation, and clustering spectrum. The properties of the sampled network turn out to be sums of quantities for the original network topology weighted by the factors stemming from the sampling. Based on our analysis, we find that the sampled network may have sampling-induced network properties that are absent in the original network, which implies the potential risk of a naive generalization of the results of the sample to the entire original network. We also consider the case, when neighboring nodes have correlated attributes to show how to generalize our formalism for such sampling bias and we get good agreement between the analytic results and the numerical simulations.

\end{abstract}

\pacs{Valid PACS appear here}
\maketitle


\section{Introduction}\label{sec:intro}

Mapping out the underlying network is an essential part of studying complex systems. Accordingly, many different networks have been constructed from empirical data sets, but this procedure is subject to noise and various biases thus being of limited applicability. 
This is especially the case when the network is huge like the internet, world wide web or human society, where unavoidably only a sample of the whole network can be analyzed that is inherently likely to cause some biases. Also the identification of 
links in a network could be the cause of bias unless all the links are equally measurable.
For example, in communication-based social networks difficulties may arise when one wants to detect social links between people using different means of communication.
The consequences of these kinds of sampling could hinder 
the generalization of the properties observed in the sample to the case of the entire system. 
For instance, the sampling bias may make the degree distribution look like a power law even when the original degree is Poissonian \cite{han2005effect,clauset2005accuracy}. Furthermore, the peaked degree distribution of social interactions is transformed to a monotonic one if only one communication channel is sampled~\cite{torok2016big}.
Also, other network quantities such as degree correlations, centrality measures, and clustering properties, could undergo nontrivial bias effects 
depending on how networks are sampled~\cite{stumpf2005subnets,stumpf2005sampling,kim2007reliability,mizutaka2018disassortativity,lee2006statistical,ahmed2014network,da2007exploring,yoon2007statistical,dall2006exploring,son2012sampling,viger2007real,eom2014generalized,eom2015tail,newman2018network}.
Thus, understanding the effect of sampling biases is crucial in interpreting empirical data better and in studying the original systems.

There have been a number of theoretical and numerical studies on network sampling since its significance was recognized.
The sampling methods studied so far are classified as 
random node sampling~\cite{stumpf2005subnets,stumpf2005sampling,kim2007reliability,mizutaka2018disassortativity,lee2006statistical,ahmed2014network},
random link sampling~\cite{newman2018network,lee2006statistical,ahmed2014network},
and path-based sampling~\cite{han2005effect,lee2006statistical,ahmed2014network,da2007exploring,yoon2007statistical,eom2014generalized,dall2006exploring,clauset2005accuracy,son2012sampling,viger2007real}.
The path-based sampling is a class of methods that sample 
nodes and links while traversing the network from certain nodes, which includes breadth first search, depth first search, snowball sampling, random walk sampling, and trace-route sampling.
For these sampling methods, the effects of the sampling biases are understood well and algorithms to improve inference of the original network properties have also been suggested~\cite{ahmed2014network,eom2014generalized,eom2015tail,viger2007real,newman2018network}.

In this paper, we study another class of network sampling, where links are sampled with a probability depending on the attributes of the nodes. We assume 
that node $i$ has an attribute $h_i$ and that the links between nodes $i$ and $j$ are sampled according to the probability depending on $h_i$ and $h_j$. This is the case when a multiplex is sampled by a limited number of layers. Usually it depends on the node attributes in which layer a link can be found. For example, in the network of social relationships a family tie between two individuals or nodes is there if both have the attribute of belonging to the same family.
This class includes the model of Ref.~\cite{torok2016big}, which was proposed to explain the commonly observed monotonic degree distribution in the data sets of social networks sampled by single communication channel.
While people communicate using various communication channels, online and offline, the sources of the data sets are often limited to a single communication channel due to technical and privacy reasons.
Thus, extracting a single communication channel is regarded as a sampling process, through which nontrivial biases are inevitably induced.
Because each person has a different tendency to select the communication channel~\cite{hargittai2015bigger} and this is adjusted to the preferences of the communication partner, the sampling process is plausibly modeled by introducing the attributes for each person rather than by random or path-based sampling methods, such that the sampling probability depends on the two communicating persons' attributes.


From a mathematical point of view, the sampling model we are going to study here is a generalization of a class of the network generation models with hidden variables~\cite{boguna2003class,caldarelli2002scale}.
In this class of models, starting from an empty network, hidden variables are assigned to the nodes, and links are generated according to a function of the hidden variables $h_i$ and $h_j$.
However, in this paper, the network is obtained by sampling from an original network having certain properties.
Hence, the sampling model studied here is equivalent to the model studied in~\cite{boguna2003class,caldarelli2002scale} when the original network is a complete graph.

This paper is organized as follows:
In Sec.~\ref{sec:model_analysis}, we present rigorous analytic forms of the degree distribution $P(k)$, degree correlation $k_{nn}(k)$, and clustering spectrum $c(k)$ for the sampled network in the case where $h$ is independently drawn 
from a certain distribution $\rho(h)$.
In Sec.~\ref{sec:examples}, we apply the results to some concrete examples.
Especially, we will investigate the model proposed in Ref.~\cite{torok2016big} and see how the original network affects the sampled network.
Then, in Sec.~\ref{sec:correlated_h}, we numerically study the case where the hidden variables of neighboring nodes in the original network are correlated with each other.
Section~\ref{sec:conclusion} is devoted to summary and discussion.

\section{Model and Analysis}\label{sec:model_analysis}

\begin{figure}[tbp]
  \begin{center}
    \includegraphics[width=\columnwidth]{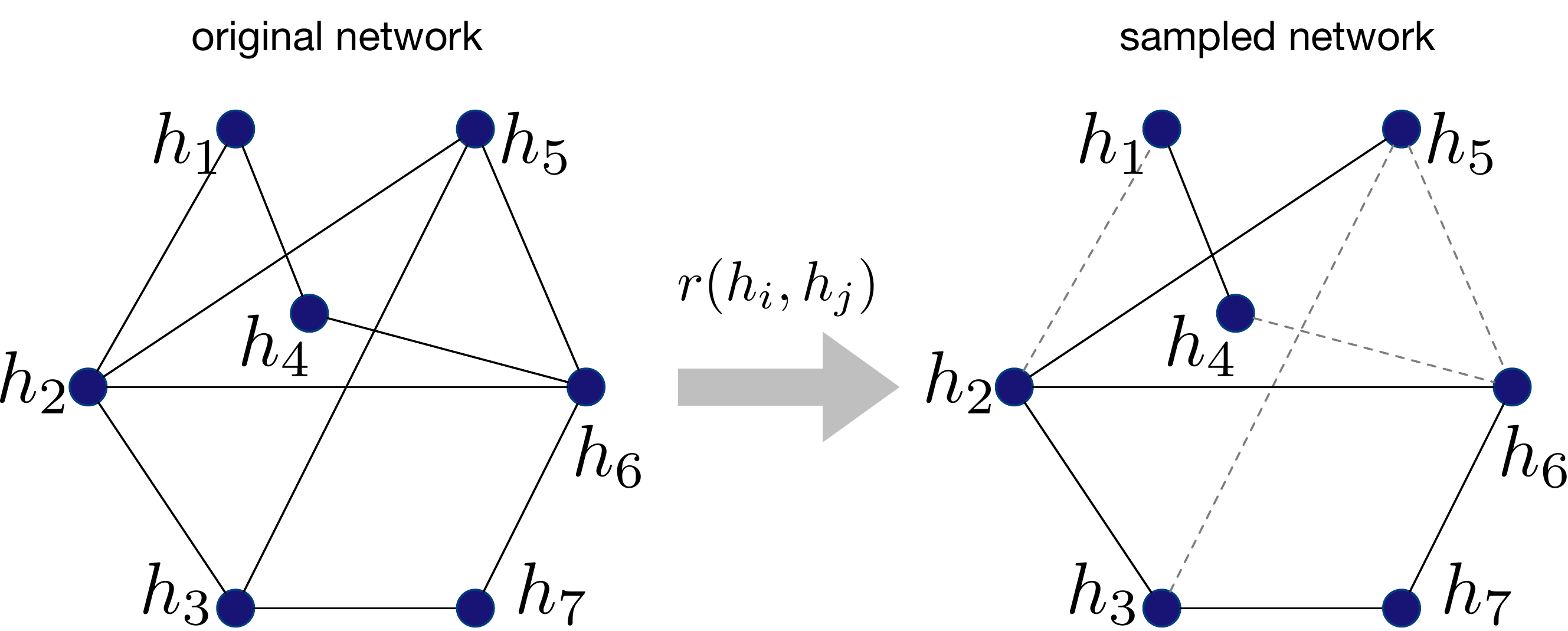}
    \caption{A schematic diagram showing the sampling method studied in this paper.
    For each node $i$, a hidden variable $h_i$ is drawn randomly from $\rho(h)$.
    A link $ij$ in the original network is sampled with a probability $r(h_i,h_j)$, where $h_i$ and $h_j$ are the hidden variables of the nodes $i$ and $j$.
    }
    \label{fig:sampling_diagram}
  \end{center}
\end{figure}

We define the model of sampling as follows (see Fig.~\ref{fig:sampling_diagram}).
First, a hidden variable $h$ is assigned to each node in the original network, where each of the hidden variables is randomly and independently drawn from the distribution $\rho(h)$.
Then, a link between nodes $i$ and $j$ is sampled with the probability $r(h_i,h_j)$, where we assume that it is a symmetric function with respect to $h_i$ and $h_j$.
Although, in this paper, we mainly consider $h$ as a scalar variable, it is straightforward to extend the model such that a node has a vector attribute (a set of attributes) $\mathbf{h}$, similarly to the Axelrod's model~\cite{axelrod1997dissemination} or the model in Ref.~\cite{papadopoulos2012popularity}, which we will discuss in Sec.~\ref{subsec:vec_attr}.

Hereafter, we denote the degree in the original network by $\kappa$, the degree distribution of the original network by $P_o(\kappa)$,
the conditional probability that a neighbor of degree-$\kappa$ node has degree $\kappa'$ by $p_o(\kappa'|\kappa)$,
and the average local clustering coefficient of degree-$\kappa$ nodes by $c_o(\kappa)$.

\subsection{Degree distribution}\label{subsec:degree}

Let us consider links around a node which has hidden variable $h$.
Since the hidden variables of the neighbors are independently given, the probability that a link around the node with $h$ is sampled is
\begin{equation}\label{eq:r_bar_h}
    \bar{r}(h) = \int dh'\rho(h')r(h,h').
\end{equation}

Since $h$ is independently given to each node, the probability distribution of the degree sampled around a node is a binomial distribution given as a function of the hidden variable $h$ and the original degree $\kappa$ of the focal node:
\begin{equation}
    g(k|h,\kappa) = \binom{\kappa}{k} \bar{r}(h)^{k} [1-\bar{r}(h)]^{\kappa-k}.
\end{equation}
This formula is not only valid for the case of independent hidden variables but 
also for a class of dependent variables with appropriately calculated $\bar{r}(h)$, which will be discussed in Sec.~\ref{sec:correlated_h}.

The degree distribution of the sampled network $P(k)$ is therefore written as
\begin{eqnarray}\label{eq:P_k}
    P(k) &=& \sum_{\kappa}\int dh g(k|h,\kappa) \rho(h) P_o(\kappa) \nonumber \\
    &=& \langle 1 \rangle_{h,\kappa},
\end{eqnarray}
where we defined the weighted sum over $h$ and $\kappa$ as
\begin{equation}\label{eq:weighted_average}
\langle X \rangle_{h,\kappa} \equiv \sum_{\kappa}\int dh g(k|h,\kappa) \rho(h) P_o(\kappa) X.
\end{equation}
Thus, the degree distribution of the sampled network depends only on $\rho(h)$, $r(h,h')$, and $P_o(\kappa)$.
It is independent of higher order correlations in the original network such as the degree correlation or clustering coefficient.

Since $g(k|h,\kappa)$ is of binomial form, the average degree in the sampled network is simply written as
\begin{eqnarray}\label{eq:avg_k}
    \langle k \rangle &=& \sum_{k}kP(k) \nonumber \\
    &=& \sum_{\kappa}\int dh \left[\sum_{k}kg(k|h,\kappa) \right] P_o(\kappa) \rho(h) \nonumber \\
    &=& \langle \kappa \rangle \bar{r},
\end{eqnarray}
where $\langle \kappa \rangle = \sum_{\kappa}\kappa P_o(\kappa)$ is the average degree in the original network, and
\begin{equation}
  \bar{r} \equiv \iint dhdh' r(h,h') \rho(h)\rho(h')  
\end{equation}
is the average sampling probability.

Similarly, one can calculate $\langle k^2 \rangle$ as
\begin{eqnarray}
    \langle k^2 \rangle &=& \sum_{k}k^2P(k) \nonumber \\
    &=& \sum_{\kappa}\int dh \left[\sum_{k}k^2g(k|h,\kappa) \right] P_o(\kappa) \rho(h) \nonumber \\
    &=& \langle\kappa\rangle\bar{r} + \left\langle \kappa(\kappa-1) \right\rangle \int dh \bar{r}(h)^2 \rho(h).
\end{eqnarray}
Thus, the second moment of the sampled degree distribution is written as a function of the first and the second moments of $P_o(\kappa)$ and the weighted average of $\bar{r}(h)^2$.
In general, the $n$th moment of $P(k)$ can be obtained as the $n$th derivatives of the characteristic function. The characteristic function is
\begin{eqnarray}
\phi(t) &=& \sum_{k} e^{ikt} P(k) \nonumber \\
  &=& \int dh \sum_{\kappa} \left[ 1-\bar{r}(h)+\bar{r}(h)e^{it} \right]^{\kappa} \rho(h) P_o(\kappa).
\end{eqnarray}
Thus, the $n$th moment of the sampled degree depends on up to the $n$th moments of $P_o(\kappa)$ and the weighted average of $\bar{r}(h)^{n}$.

\subsection{Degree correlation}\label{subsec:knn}

The degree correlation between neighboring nodes in the sampled network can be characterized by the conditional distribution as
\begin{widetext}
\begin{equation}
P(k'|k) = \iint dh dh'\sum_{\kappa, \kappa'} g(k'-1|h',\kappa'-1) p(h',\kappa'|h,\kappa) g^{\ast}(h,\kappa|k),
\end{equation}
\end{widetext}
where $g^{\ast}(h,\kappa|k)$ is the joint conditional probability of $h$ and $\kappa$ given $k$.
Since one connection has already been used up for the conditional edge with $h$, $g(k'-1|h',\kappa'-1)$ gives the probability that a node with $(h',\kappa')$ ends up with degree $k'$.
Using the Bayes' formula, we obtain
\begin{equation}
    g^{\ast}(h,\kappa|k) = \frac{1}{P(k)} \rho(h) P_{o}(\kappa) g(k|h,\kappa).
\end{equation}
The conditional probability $p(h',\kappa'|h,\kappa)$ is the probability that a neighbor of a node with $(h,\kappa)$ in the sampled network has the hidden variable $h'$ and the original degree $\kappa'$.
Since $h$ is assigned independently to nodes, it is written as the product of two factors:
\begin{eqnarray}\label{eq:h_kappa_corr}
p(h',\kappa'|h,\kappa)
 &=& p(h'|h) p_o(\kappa'|\kappa),
\end{eqnarray}
where the conditional probability $p(h'|h)$ is written as
\begin{equation}\label{eq:p_hprime_given_h}
p(h'|h) = \frac{r(h',h)\rho(h')}{\bar{r}(h)}.
\end{equation}
Note that there is a correlation of $h$ between neighboring nodes after the sampling even though there is no correlation in the original network.
Sampling induces correlations of neighboring $h$ values.

The degree correlation is then given by
\begin{widetext}
\begin{equation}
    P(k'|k) = \frac{1}{P(k)} \iint dhdh' \sum_{\kappa, \kappa'} g(k'-1|h',\kappa'-1) p(h'|h) p_o(\kappa'|\kappa) g(k|h,\kappa) \rho(h) P_{o}(\kappa).
\end{equation}
Using this, the average degree of neighbors of a degree-$k$ node is
\begin{eqnarray}\label{eq:k_nn}
k_{nn}(k) &=& \sum_{k'} k' P(k'|k) \nonumber \\
  &=& \sum_{k'}\iint dhdh' \sum_{\kappa,\kappa'} k'g(k'-1|h',\kappa'-1) p(h'|h) p_o(\kappa'|\kappa) \frac{g(k|h,\kappa)\rho(h)P_{o}(\kappa)}{P(k)} \nonumber \\
  &=& 1 + \int dh\sum_{\kappa} \frac{g(k|h,\kappa)}{P(k)} \rho(h)P_{o}(\kappa) \left[\int dh'\bar{r}(h')p(h'|h) \right]  \left[ \sum_{\kappa'} (\kappa'-1) p_o(\kappa'|\kappa) \right] \nonumber \\
  &=& 1 + \frac{ \left\langle \bar{r}_{nn}(h) \left(\kappa_{nn}(\kappa)-1\right) \right\rangle_{h,\kappa} }{P(k)},
\end{eqnarray}
\end{widetext}
where $\bar{r}_{nn}(h)$ is the average sampling probability of the links around a neighbor of a node having $h$ and $\kappa_{nn}(\kappa)$ is the average neighbor degree of a degree-$\kappa$ node in the original network.
Each of these is defined as
\begin{equation}
  \bar{r}_{nn}(h) = \int dh' \bar{r}(h')p(h'|h)
\end{equation}
and
\begin{equation}
  \kappa_{nn}(\kappa) = \sum_{\kappa'}\kappa'p_o(\kappa'|\kappa).
\end{equation}
Thus, $k_{nn}(k)$ is written as a weighted sum of $\bar{r}_{nn}(h)\left(\kappa_{nn}(\kappa)-1\right)$, which is the product of the correlations of hidden variables and the original degrees.
It depends on $P_o(\kappa)$ and $\kappa_{nn}(\kappa)$ but is independent of other higher order correlations.

In general, the hidden variables of neighboring nodes in the sampled network show a correlation even if $h$ is originally independent of the neighbors. This implies that the correlation between $h$s in the sampled network is induced by the sampling.
Here, we note that the correlation of hidden variables in the sampled network $p(h'|h)$ is totally independent of the original network because it depends only on the functional forms of $\rho(h)$ and $r(h,h')$.
The hidden variable averaged over neighbors of a node having $h$ is written as
\begin{eqnarray}\label{eq:h_nn}
  h_{nn}(h) &=& \int dh' h' p(h'|h)
\end{eqnarray}
using Eq.~(\ref{eq:p_hprime_given_h}).

The sampling induced correlations in $h$ disappear when $r(h_i,h_j)$ is factorized such that
$r(h_i,h_j)=r'(h_i)r'(h_j)$ irrespective of the functional form of $\rho(h)$.
This is because the conditional probability $p(h'|h)$ is independent of $h$ as follows:
\begin{eqnarray}
p(h'|h) &=& \frac{r(h,h')\rho(h')}{\int dh'' r(h,h'')\rho(h'')} \nonumber \\
  &=& \frac{r'(h')\rho(h')}{\int dh'' r'(h'')\rho(h'')}.
\end{eqnarray}

\subsection{Clustering coefficient}\label{subsec:ck}

Consider a node with hidden variable $h$ and original degree $\kappa$. In the original network, the local clustering coefficient, denoted by $c_o(\kappa)$, denotes the fraction of the pairs of neighbors having links between them.
Therefore, the local clustering coefficient of this node in the sampled network $c_{h,\kappa}$ is
\begin{eqnarray}
    c_{h,\kappa} &=& \iint dh'dh'' c_o(\kappa) r(h',h'') p(h'|h) p(h''|h) \nonumber \\
      &=& c_o(\kappa) c_h(h),
\end{eqnarray}
where
\begin{equation}
c_h(h) \equiv \int dh' dh'' r(h',h'') p(h'|h) p(h''|h).
\end{equation}
The average local clustering coefficient of a node with sampled degree $k$, denoted by $c(k)$, is given by the average of $c_{h,\kappa}$ weighed by the probability that the node has the hidden variable $h$ and the original degree $\kappa$:
\begin{eqnarray}\label{eq:c_k}
c(k) &=& \int dh \sum_{\kappa} g^{\ast}(h,\kappa|k) c_{h,\kappa} \nonumber \\
  &=& \frac{1}{P(k)} \int dh \sum_{\kappa} g(k|h,\kappa) \rho(h) P_o(\kappa) c_o(\kappa) c_{h}(h) \nonumber \\
  &=& \frac{ \left\langle c_o(\kappa) c_h(h) \right\rangle_{h,\kappa} }{P(k)}.
\end{eqnarray}
Therefore, $c(k)$ is given by the weighted sum of the product of $c_o(\kappa)$ and $c_h(h)$.
It is notable that it does not depend on the degree correlation between neighbors in the original network. It depends only on $P_o(\kappa)$ and $c_o(\kappa)$.
The average clustering coefficient in the sampled network is then given as
\begin{equation}
    \langle c \rangle = \sum_{k=2}^{\infty}c(k)P(k).
\end{equation}

The equations for the sampled network properties are summarized in Table~\ref{tab:eq_summary}.

\begin{table*}[t]
    \centering
    \begin{tabular}{c|c|c|c}
    \hline
        Network property & Analytic form & Equation & Dependency on the original network \\
        \hline
        $\langle k \rangle$ & $\langle\kappa\rangle\bar{r}$ & (\ref{eq:avg_k}) & $\langle\kappa\rangle$ \\
        $P(k)$ & $\langle 1 \rangle_{h,\kappa}$ & (\ref{eq:P_k}) & $P_o(\kappa)$ \\
        $k_{nn}(k)$ & $1 + \frac{\left\langle \bar{r}_{nn}(h)\left(\kappa_{nn}(\kappa)-1\right) \right\rangle_{h,\kappa} }{P(k)}$ & (\ref{eq:k_nn}) & $P_o(\kappa)$, $\kappa_{nn}(\kappa)$ \\
        $c(k)$ & $\frac{\left\langle c_o(\kappa)c_h(h) \right\rangle_{h,\kappa}}{P(k)}$ & (\ref{eq:c_k}) & $P_o(\kappa)$, $c_o(\kappa)$\\
        $h_{nn}(h)$ & $\int dh' h' p(h'|h)$ & (\ref{eq:h_nn}) & None 
        \\
    \hline
    \end{tabular}
    \caption{
    Summary of the analytic equations for the sampled network properties.
    On the right column, the dependency on the original network properties is shown.
    For instance, the sampled degree distribution depends only on the original degree distribution $P_o(\kappa)$.
    The weighted sum $\langle X \rangle_{h,\kappa}$ is defined by Eq.~(\ref{eq:weighted_average}).
    When neighboring $h$s are correlated, one can refer to Eq.~(\ref{eq:r_bar_h_corr}) and Eq.~(\ref{eq:p_h_given_h_corr}) instead of Eq.~(\ref{eq:r_bar_h}) and Eq.~(\ref{eq:p_hprime_given_h}), respectively.
    }
    \label{tab:eq_summary}
\end{table*}

\subsection{Vector attribute}\label{subsec:vec_attr}

Although we have assumed that a hidden variable $h$ for a node is a scalar, it is straightforward to extend our analysis for general attributes.
In general, a node may have a vector attribute $\mathbf{h}$ of dimension $d$ whose elements may be continuous or have discrete numbers, $\mathbb{R}^{d}$.
The probability density function $\rho(\mathbf{h}): \mathbb{R}^d \mapsto \mathbb{R}_{+}$ and the sampling probability $r(\mathbf{h}, \mathbf{h'}): \mathbb{R}^d \times \mathbb{R}^d \mapsto \mathbb{R}_{+}$
are defined in the extended spaces.

The analytic equations shown in the previous subsections are valid by replacing integrals over a scalar $h$ by the integrals over the vector $\mathbf{h}$.
For instance, Eq.~(\ref{eq:r_bar_h}) is
\begin{equation}
    \bar{r}(\mathbf{h}) \equiv \int d\mathbf{h'} \rho(\mathbf{h'})r(\mathbf{h},\mathbf{h'}).
\end{equation}
Similarly, Eq.~(\ref{eq:weighted_average}) is redefined as
\begin{equation}
    \langle X \rangle_{\mathbf{h},\kappa} \equiv \sum_{\kappa}\int d\mathbf{h} g(k|\mathbf{h},\kappa) \rho(\mathbf{h}) P_o(\kappa) X.
\end{equation}
In the case when an element of $\mathbf{h}$ is a discrete value, the corresponding integral is replaced by a summation.
After these modification, the other equations in the previous subsections are still valid.

\section{Examples}\label{sec:examples}

\subsection{Sampling with a generalized mean of hidden variables}\label{subsec:genmean}

In this section, we study some concrete examples of $\rho(h)$ and $r(h_i,h_j)$.
The first model we are going to study is the sampling method proposed in Ref.~\cite{torok2016big}, which was introduced to explain monotonically decreasing degree distributions, as they are commonly observed in various data sets of social networks.
Such monotonically decreasing degree distribution, indicating that the most frequent degree is one, is considered to be an outcome of the bias of sampling a single communication channel.
With reference to our everyday experience we can consider it very unlikely to find a randomly selected person who has only one friend or social tie,
indicating that the original social network should have a degree distribution with a peak at a degree larger than one.
Canonical sampling methods, such as random node or link sampling or snowball sampling, are not suitable for explaining this discrepancy since they may result in 
the sampled degree distribution that is not monotonically decreasing.
In contrast, the model presented in Ref.~\cite{torok2016big} can be considered simple and plausible in explaining the monotonically decreasing degree distribution in the sample network. It assumes that the monotonically decreasing degree distribution is attributed to the mixture of the rare and frequent users of the communication service.
In the model, the tendency for a person to use the communication channel is represented by a nodal attribute $h$ and the link sampling probability is related to the channel selection. 

The model is defined such that each node has a scalar value $h$, which is independently drawn from the distribution $\rho(h)$. The value of $h$ denotes how much one person favors an online service or a communication channel.
The distribution of the hidden variables $\rho(h)$ is a Weibull distribution truncated at $h=1$:
\begin{equation}\label{eq:weibull_rho_h}
    \rho(h) =
    \begin{cases}
    c \frac{\alpha}{h_0} \left(\frac{h}{h_0}\right)^{\alpha-1} \exp{\left[-\left(\frac{h}{h_0}\right)^{\alpha}\right]} & \text{when $0 \leq h \leq 1$} \\
    0 & \text{otherwise},
    \end{cases}
\end{equation}
where $c = \left[1-e^{-(1/h_0)^{\alpha}}\right]^{-1}$ is a normalization constant.

As for the sampling probability we consider the generalized mean of the two hidden variables:
\begin{equation}\label{eq:genmean_r}
    r(h,h') =
    \begin{cases}
    \left( \frac{h^{\beta}+h'^{\beta}}{2} \right)^{1/\beta}  & \text{when $\beta \neq 0$} \\
    \sqrt{h h'}                                              & \text{when $\beta = 0$},
    \end{cases}
\end{equation}
where $\beta$ is an exponent characterizing the generalized mean which takes the form of arithmetic, geometric, or harmonic mean for $\beta=1$, $0$, or $-1$, respectively.
In the limits of $\beta \to \infty$ and $\beta \to -\infty$, the sampling probabilities are equivalent to $\max\{h,h'\}$ and $\min\{h,h'\}$, respectively.

\begin{figure}[htbp]
  \begin{center}
    \includegraphics[width=\columnwidth]{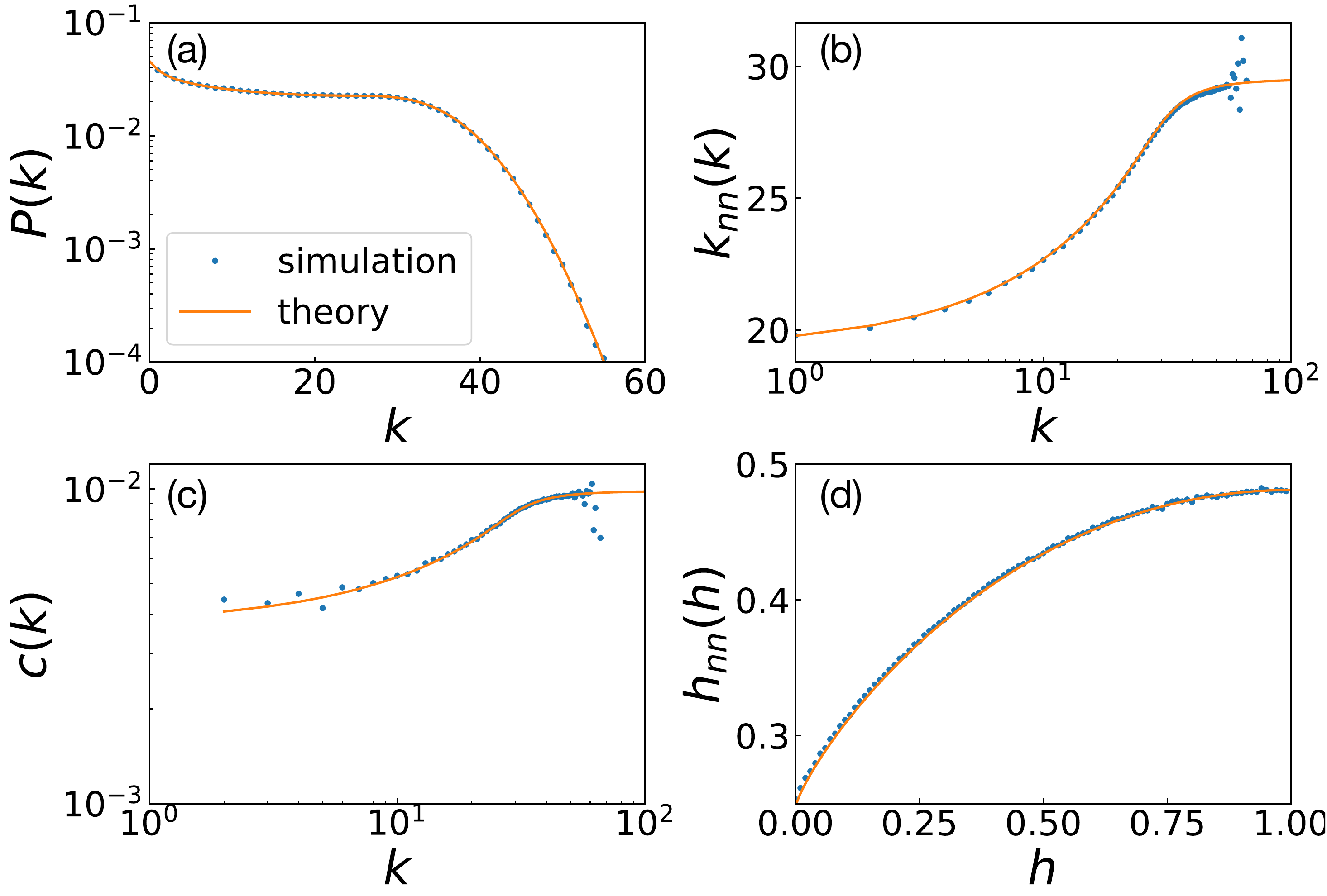}
    \caption{(a) Degree distribution $P(k)$, (b) degree correlation $k_{nn}(k)$, (c) clustering coefficient $c(k)$, (d) average neighbor hidden variable $h_{nn}(h)$ of the sampled networks.
    Simulation results are compared with theoretical equations in Table~\ref{tab:eq_summary}.
    The network is sampled from the Erd\"{o}s-R\'{e}nyi random graph of size $N=5000$ and the link density $p=0.03$.
    The parameter values for sampling are $\alpha=0.8$, $h_0=0.3$, and $\beta=-\infty$.
    Under this setting, $12\%$ of the links are sampled. Simulation results are averaged over $200$ independent runs.
    }
    \label{fig:compare_sim_eq}
  \end{center}
\end{figure}

As a demonstration, we compare our analytic expressions in Sec.~\ref{sec:model_analysis} with the Monte Carlo simulations.
As an original network, we use an Erd\"{o}s-R\'{e}nyi random graph of size $N=5000$ and link density $p=0.03$.
The parameters for the sampling are $h_0=0.3$, $\alpha=0.8$, and $\beta=-\infty$.
Figure~\ref{fig:compare_sim_eq} shows that our analytic results are in excellent agreement with the results from Monte Carlo simulations, demonstrating 
the validity of our analysis 
in the previous Section.

As shown numerically in Section V of Ref.~\cite{torok2016big}, the sign of $\beta$ is crucial to obtain a monotonically decreasing degree distribution from an original network having a mode at higher than one.
For all the results showing monotonically decreasing degree distributions, $\beta \leq 0$ is satisfied.
This is reasonable because the sampling probability around a node with $h$, $\bar{r}(h)$, does not go to $0$ for a positive $\beta$ even when $h \to 0$.
In other words, even with an infinitesimal $h$, a node has a positive finite sampling probability of links, which makes the low-degree nodes rare and yields the peaked degree distribution.

Although a negative $\beta$ reproduces a monotonically decreasing degree distribution, it also causes side-effects in other network statistics. 
The degree correlation $k_{nn}(k)$ shows an increasing behavior while the original network does not have any degree correlation between neighbors, 
strongly implying that degree assortativity is induced by the sampling.
The sign of $\beta$ plays a pivotal role for the sampling-induced assortativity.
In Ref.~\cite{torok2016big}, it was shown that the neighboring $h$ in the sampled network is positively (or negatively) correlated for a negative (or positive) $\beta$ regardless of the original network topology or the functional form of $\rho(h)$.
The correlation of $h$ induced by the sampling also causes the correlation of $k$ in the sampled network. As a matter of fact, 
an increasing $k_{nn}(k)$ is commonly observed in various empirical data sets of social networks.
A plausible explanation for degree assortativity could be the homophily mechanism, however, the model implies that it may be an outcome of the bias caused by sampling a single communication channel. 
We cannot naively conclude that the original network shows a positive correlation even if the data set shows a positive degree correlation between neighbors. The fact that high degree nodes are likely to be connected due to homophily is not enough for degree assortativity as it is easy to construct disassortative networks, with interconnected hubs~\cite{colizza2006detecting}.

The local clustering coefficient as a function of the degree $c(k)$ is also biased. As seen in Fig.~\ref{fig:compare_sim_eq}(c), it shows an increasing behavior while it originally had a flat profile.
This increasing behavior is also explained by the sampling induced assortativity.
The low $k$ nodes mostly consist of nodes having low $h$.
Because of the positive correlation in $h$, the neighboring nodes around a low-degree node also tends to have a low $h$.
Therefore, the probability of making a link between neighbors is low.
However, for a high-degree node, the hidden variables of its neighbors tend to be high as well, yielding a higher local clustering coefficient.
As a result, $c(k)$ shows an increasing trend.
In empirical networks, however, decreasing $c(k)$ is commonly found in many data sets~\cite{ugander2011anatomy,szell2010measuring,onnela2007analysis,jo2018stylized}.
This is an unrealistic aspect of the model, which might originate from oversimplification of the sampling process or due to  mechanisms other than sampling.

If we increase or decrease the number of sampled links by changing $h_0$ and/or $\alpha$, then the property of the sampled network changes.
For instance, the shape of $P(k)$ becomes more similar to the original one if the fraction of sampled links gets closer to one.
However, the increasing behavior of $k_{nn}(k)$ and $c(k)$ are more robust: They remain qualitatively same as long as the sign of $\beta$ is negative.
Conversely, decreasing behaviors are found when $\beta$ is positive, indicating that the sign of $\beta$ is a crucial factor for how the sampling causes bias on $k_{nn}(k)$ and $c(k)$.

So far, we have used an Erd\"{o}s-R\'{e}nyi random graph for the original network.
One can ask how would the properties of the original network affect these results?
As we have shown in the previous section, the degree correlation in the sampled network $k_{nn}(k)$ consists of the contributions both from the original degree correlation $\kappa_{nn}(\kappa)$ and the correlation of $h$, $\bar{r}_{nn}(h)$.
To see how it depends on $\kappa_{nn}(\kappa)$, we conducted a thought experiment by manually assigning an increasing or a decreasing function to $\kappa_{nn}(\kappa)$.
For simplicity, we adopt linearly increasing or decreasing functions, $\kappa_{nn}(\kappa) = \pm 0.2\left(\kappa-\langle \kappa \rangle\right) + \langle\kappa_{nn}\rangle$, where $\langle \kappa \rangle = \sum_{\kappa}\kappa P_o(\kappa)$ and $\langle \kappa_{nn} \rangle = \sum_{\kappa}\kappa_{nn}(\kappa)P_o(\kappa)$.
We can calculate how $k_{nn}(k)$ would look like if such assortative and disassortative networks existed using the equations in Table~\ref{tab:eq_summary}.

\begin{figure}[tbp]
  \begin{center}
  \includegraphics[width=\columnwidth]{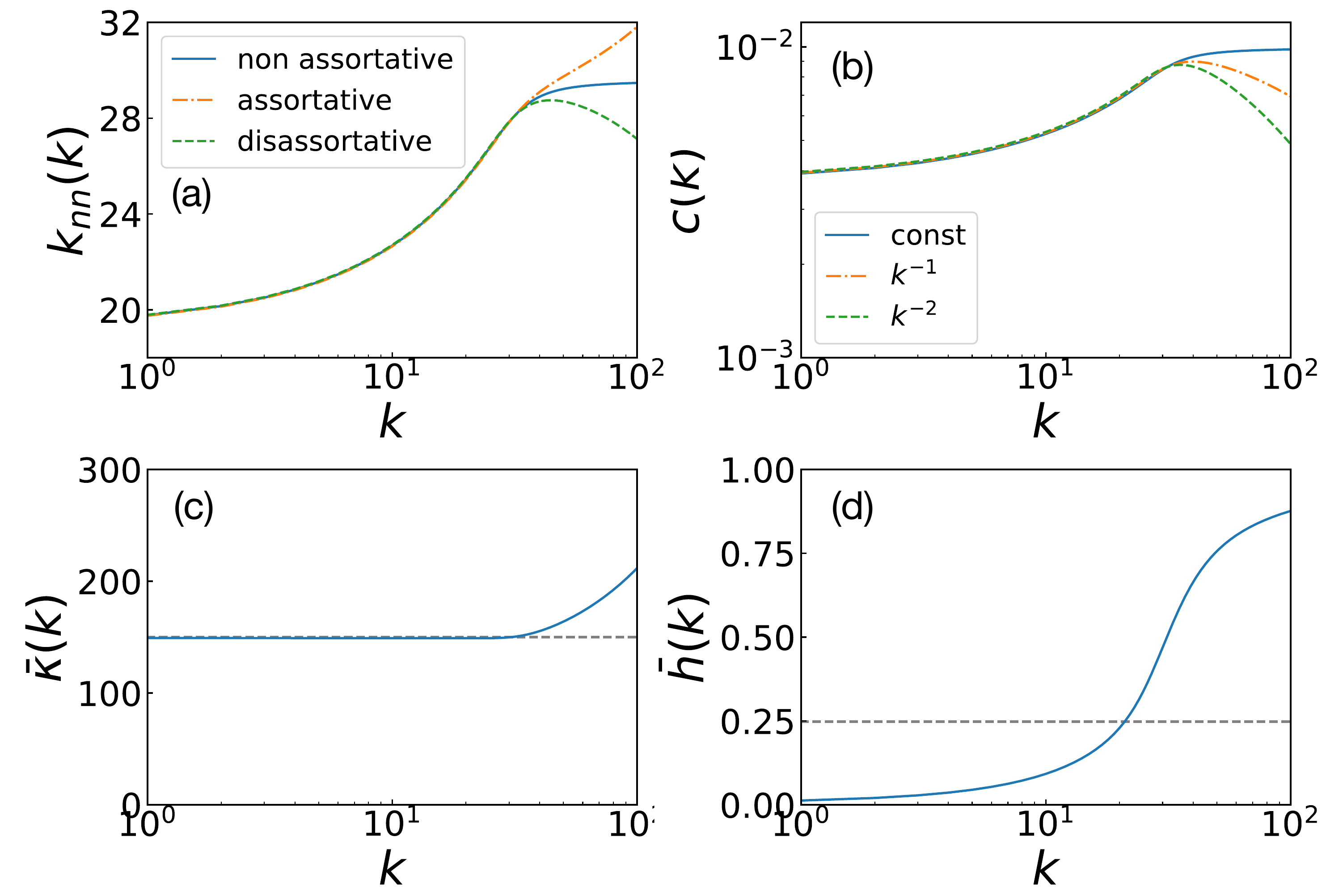}
    \caption{(a) The degree correlation $k_{nn}(k)$ when the original network is nonssortative, assortative, and disassortative.
    (b) The clustering spectrum $c(k)$ of the sampled network when the original clustering spectrum is $c_o(\kappa) \propto \kappa^{0} \text{(const)}$, $\kappa^{-1}$, and $\kappa^{-2}$.
    (c) The average original degree given the sampled degree, $\bar{\kappa}(k)$.
    The inferred original degree $\bar{\kappa}(k)$ 
    is just slightly below the original average degree $\langle\kappa\rangle$, depicted as a horizontal dashed line, for $k < k^{\ast}$.
    (d) The average hidden variable given sampled degree $\bar{h}(k)$.
    The horizontal dashed line depicts $\langle h \rangle = \int dh h \rho(h)$.
    The other settings are the same as described in the caption of Fig.~\ref{fig:compare_sim_eq}.
    }
    \label{fig:thought_experiment}
  \end{center}
\end{figure}

The result of the thought experiment is shown in Fig.~\ref{fig:thought_experiment}(a).
The distribution $\rho(h)$ and the sampling probability $r(h,h')$ are kept same as the previous one.
The figure shows that the curves almost collapse into a single curve when $k < k^{\ast}$, and a significant difference appears after $k$ exceeds $k^{\ast}$, where $k^{\ast} \approx 30$.
It indicates that $k_{nn}(k)$ for $k < k^{\ast}$ mostly reflects the property of the sampling rather than the original network property.
Because the majority of the nodes in the sampled network have a degree smaller than $k^{\ast}$ [see Fig.~\ref{fig:compare_sim_eq}(a)], it is practically difficult to obtain the information of the original network from the sampled network.

A similar experiment is conducted for $c(k)$.
We calculate $c(k)$ by assuming three possible cases: $c_o(\kappa) \propto \kappa^{0}$, $\kappa^{-1}$, and $\kappa^{-2}$. For a fair comparison, we keep the average clustering coefficient in the original network $\langle c_o \rangle = \sum_{\kappa} P_o(\kappa)c_o(\kappa)$ constant.
The results are shown in Fig.~\ref{fig:thought_experiment}(b).
Similar to $k_{nn}(k)$, the clustering spectrum $c(k)$ shows the dependency on the original network only for $k > k^{\ast}$.
The low-degree behavior is determined by the dependency on $h$ hence it does not contain much information about the original network property.

These results of the thought experiment are explained by calculating the conditional probability distribution of the original degree $\kappa$ given a sampled degree $k$:
\begin{equation}\label{eq:p_kappa_given_k}
    P^{\ast}(\kappa|k) = \int dh g^{\ast}(h,\kappa | k) = \frac{P_o(\kappa)\int dh \rho(h)g(k|h,\kappa)}{\sum_\kappa P_o(\kappa)\int dh \rho(h)g(k|h,\kappa)}.
\end{equation}
If $P^{\ast}(\kappa|k)$ is identical to $P_o(\kappa)$, then the network property around a degree-$k$ node is determined only by $h$ dependency irrespective of $\kappa$.
In other words, the difference between $P^{\ast}(\kappa|k)$ and $P_o(\kappa)$ serves as an indicator of the relevance of the original network properties.
We calculated the expected original degree conditioned on a sampled degree, which is defined as $\bar{\kappa}(k) = \sum_{\kappa}\kappa P^{\ast}(\kappa|k)$.
As shown in Fig.~\ref{fig:thought_experiment}(c), $\bar{\kappa}(k)$ remains constant at $\langle \kappa \rangle$ for $k < k^{\ast}$ while it shows deviations only when $k > k^{\ast}$, supporting the observations so far.
We also calculated the Kullback-Leibler (KL) divergence between these two distributions for a given $k$, which is defined by
\begin{equation}
    D(k) = \sum_{\kappa} P^{\ast}(\kappa|k) \ln{\left[\frac{P^{\ast}(\kappa|k)}{P_o(\kappa)}\right]}.
\end{equation}
The result (not shown) remains near zero for $k<k^{\ast}$, which is again consistent with the behaviors in Figs.~\ref{fig:thought_experiment}(a) and \ref{fig:thought_experiment}(b).

Since the dependence on $\kappa$ is not significant, the sampled network topology can mostly be attributed to the dependence on $h$.
The conditional probability distribution of $h$ given the sampled degree $k$ which is a counterpart of Eq.~(\ref{eq:p_kappa_given_k}), is written as
\begin{equation}
    P^{\ast}(h|k) = \sum_{\kappa} g^{\ast}(h,\kappa | k).
\end{equation}
The expected value $\bar{h}(k) = \int dh h P^{\ast}(h|k)$ is shown in Fig.~\ref{fig:thought_experiment}(d).
In contrast to Fig.~\ref{fig:thought_experiment}(c), $\bar{h}(k)$ is significantly different from the original average $\langle h \rangle$ for all range of $k$.
Thus, the sampled network reflects $h$ of each node while the contribution of the original network topology is marginal.

The transition point $k^{\ast}$ is 
estimated by the expected $k$ for a node having the maximum value of $h$.
When $h$ is maximum ($h = 1$ in this case), the expected degree after sampling is $\langle k \rangle_{h=1} = \langle \kappa \rangle \bar{r}(1) \approx 37$, which agrees well with the numerical value of $k^{\ast}$.
While $k$ is determined both by $\kappa$ and $h$ in general, it is mostly determined by $h$ for the nodes $k < \langle k \rangle_{h=1}$ as we have seen in Figs.~\ref{fig:thought_experiment}(c) and \ref{fig:thought_experiment}(d).
However, for $k > \langle k \rangle_{h=1}$, the dependence on $h$ is limited since these nodes are expected to have similar values $h \sim 1$, which makes the dependence on $\kappa$ more apparent in $k$.
Thus, we find a similarity in $k$-dependence and $\kappa$-dependence on the network properties for $k > \langle k \rangle_{h=1}$.
Similar argument should apply also to the lower bound: When $k < \langle k \rangle_{h=0}$, the original network property appears more in the sampled network although it is not visible because $\langle k \rangle_{h=0} = 0$ when $\beta \leq 0$.

In general, the property of the original network is not reflected for all $k$ but for a limited range of $k$ as this example illustrates.
It can be hard to obtain the information about the original network since the hidden variables can be a definitive factor for the sampled network topology. 
The conditional probability distributions $P^{\ast}(\kappa|k)$ or the KL divergence $D(k)$ serves as an indicator of the dependency on the original network.
It is noteworthy that $P^{\ast}(\kappa|k)$ and $D(k)$ are dependent only on $\rho(h)$, $r(h,h')$ and $P_o(\kappa)$, hence it is independent of any higher-order correlation in the original network.

The dependency of $k_{nn}(k)$ and $c(k)$ on the original network may change when the original degree distribution changes. As a trivial example, let us consider the case where $P_o(\kappa)$ is a delta function as in the case of a regular random network. The $k$-dependency of $k_{nn}(k)$ and $c(k)$ are fully attributed to the sampling hence it does not contain any information about the original network properties.
However, when the variance of $P_o(\kappa)$ is large, more information of the original network properties are likely to be reflected to the sampled networks.

\begin{figure}[tbp]
  \begin{center}
    \includegraphics[width=\columnwidth]{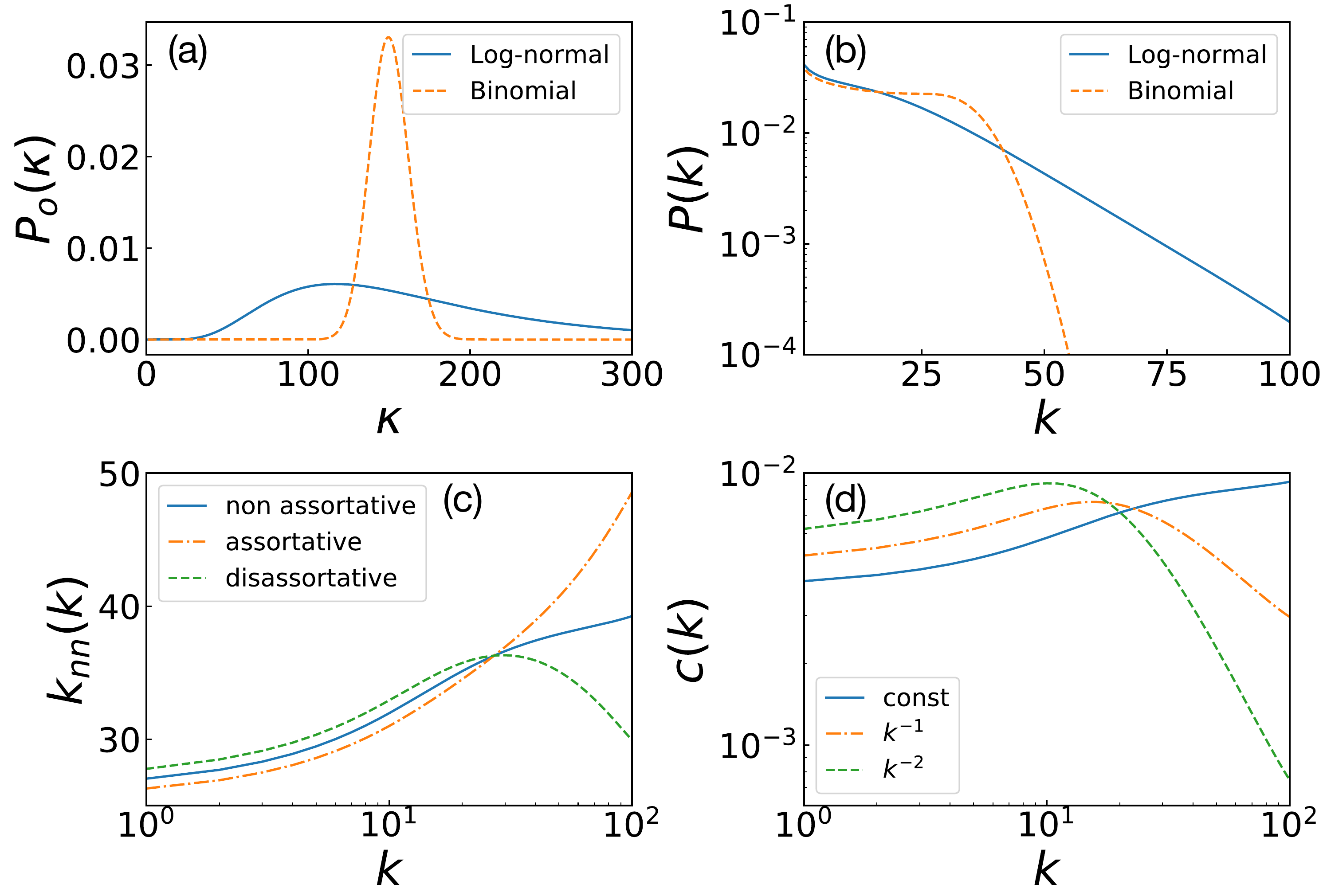}
    \caption{
    (a) The original degree distribution $P_o(\kappa)$ for the case of Binomial distribution (ER random graph) and the log-normal distribution. For the log-normal distribution, $\mu=150$ and $\sigma=0.5$ are used.
    (b) The sampled degree distributions $P(k)$ when $P_o(\kappa)$ are the log-normal and the binomial distributions. The parameters for the sampling are the same as described in the caption of Fig.~(\ref{fig:compare_sim_eq}).
    (c) The average neighbor degree $k_{nn}(k)$ for the log-normal $P_o(\kappa)$.
    The same setting is used as in Fig.~\ref{fig:thought_experiment}(a), except for $P_o(\kappa)$.
    (c) The clustering spectrum $c(k)$ for the log-normal $P_o(\kappa)$.
    The same setting is used as in Fig.~\ref{fig:thought_experiment}(b), except for $P_o(\kappa)$.
    }
    \label{fig:lognormal_pkappa}
  \end{center}
\end{figure}

We conducted the same analysis for a case where the original degree distribution is a log-normal distribution:
\begin{equation}
    P_o(\kappa) = \frac{1}{\sqrt{2\pi}\sigma \kappa}\exp{\left[ - \frac{\left(\ln{\kappa
    }-\ln{\mu}\right)^2}{2\sigma^2} \right]}.
\end{equation}
The original degree distributes more widely than the case for the Erd\H os-R\'enyi random graph as shown in Fig.~\ref{fig:lognormal_pkappa}(a).
When $P_o(\kappa)$ distributes more widely, more properties of the original network are found in the sampled networks.
Compare Figs.~\ref{fig:lognormal_pkappa}(c) and \ref{fig:lognormal_pkappa}(d) with Figs.~\ref{fig:thought_experiment}(a) and \ref{fig:thought_experiment}(b), respectively.
Although the overall profile is similar, the dependency on the original network property appears more strongly for this case.
Furthermore, the sampled degree distribution $P(k)$ for the log-normal case have a heavier tail.
The original network property is more visible in the sampled network as more data points appear in $k > k^{\ast}$.

Although the results shown above are for an extreme case $\beta = -\infty$, qualitatively similar results are found for $\beta<0$.
When $\beta$ is negative, increasing behavior is found in $k_{nn}(k)$ and $c(k)$ even when the original network shows a flat profile.
Conversely, decreasing behavior is found for positive $\beta$.
Some of the examples for other values of $\beta$ are shown in Ref.~\cite{torok2016big}.

\subsection{Sampling by a vector of hidden variables}\label{subsec:vec_attr_example}

We demonstrate another example where the hidden variable of a node is not a continuous scalar value but a vector $\mathbf{h}_{i}$.
Inspired by the Axelrod's model for the dissemination of culture~\cite{axelrod1997dissemination}, $\mathbf{h}_i$ is assumed to be a $F$-dimensional vector whose components take one of $q$ discrete values $[0,q-1]$.
Although this is nothing but a toy model, it is introduced to show the validity of the theory.

In the following, $F=2$ and $q=4$ are used, and its first and second components are denoted as $\sigma$ and $\tau$, that is, $\mathbf{h}_i = \left(\sigma_i,\tau_i \right)$.
The probability mass function $\rho(\mathbf{h})$ is the uniform distribution: $\rho(\mathbf{h}) = 1/q^2$ for any $\sigma$ and $\tau$.
For the sampling probability function $r(\mathbf{h}_i,\mathbf{h}_j)$, the following function is used:
\begin{equation}\label{eq:vector_r}
    r(\mathbf{h}_i,\mathbf{h}_j) = \frac{ \max\left\{\sigma_i,\sigma_j\right\} + c \min\left\{\tau_i,\tau_j\right\} }{(c+1)(q-1)},
\end{equation}
where $c>0$ is a parameter controlling the relative weight between the first and second terms. In this example, we use $c=2$.
As an original network, we use an Erd\"{o}s-R\'{e}nyi random graph with $N=50000$ and $p=0.003$.

\begin{figure}[tbp]
  \begin{center}
    \includegraphics[width=\columnwidth]{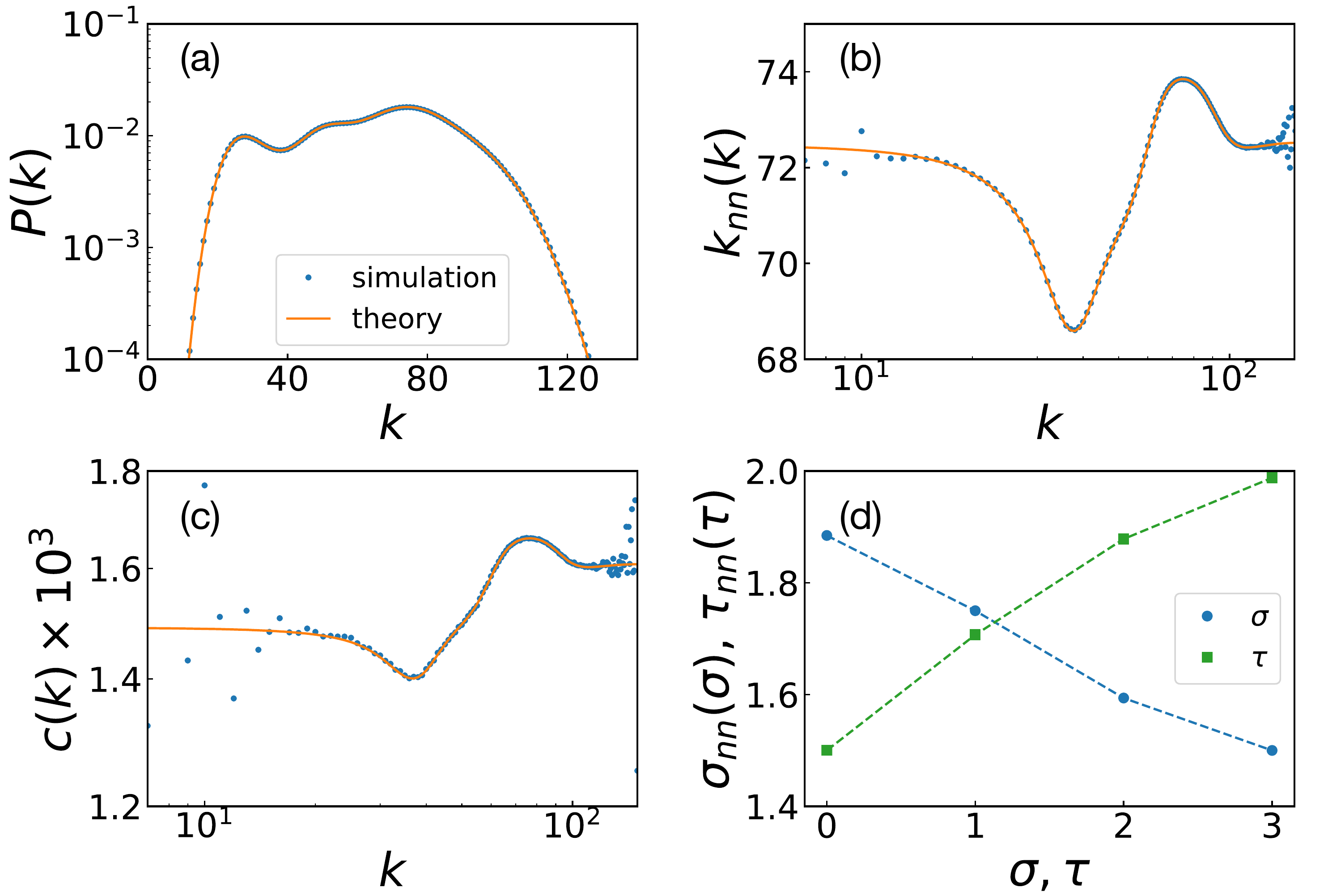}
    \caption{
    (a) Degree distribution $P(k)$, (b) average neighbor degrees $k_{nn}(k)$, (c) clustering spectrum $c(k)$ of the sampled network for the model having vector nodal attributes $\mathbf{h}$.
    Simulation results are compared with the theoretically predicted lines.
    The network is sampled with probability given by Eq.~(\ref{eq:vector_r}) with $F=2$, $q=4$ and $c=2$ from an original network of Erd\"{o}s-R\'{e}nyi random graph with $N=50000$ and $p=0.003$.
    Under this setting, approximately $43\%$ of the links are sampled.
    The simulations results are averaged over $1000$ independent runs.
    (d) Correlation of the hidden variables between neighbors in the sampled networks.
    Symbols denote the simulation results while dashed curves are theoretically predicted results.
    }
    \label{fig:compare_sim_eq_vec}
  \end{center}
\end{figure}

Figure~\ref{fig:compare_sim_eq_vec} shows the simulation results as well as theoretical prediction.
The degree distribution $P(k)$, average neighbor degree $k_{nn}(k)$, and clustering spectrum $c(k)$ are shown in Figs.~\ref{fig:compare_sim_eq_vec}(a)--\ref{fig:compare_sim_eq_vec}(c), respectively.
Although the profiles of these curves are not as simple as those in the previous Subsection, the theoretical curves perfectly coincide with the simulation results, proving the validity of our analytic approach. 
We also studied the correlation between neighboring $\sigma$ and $\tau$ by measuring the curves for $\bar{\sigma}_{nn}(\sigma)$ and $\bar{\tau}_{nn}(\tau)$, which are defined as the average of neighbors' hidden variables around a node having $\sigma$ and $\tau$, respectively.
As shown in Fig.~\ref{fig:compare_sim_eq_vec}(d), $\sigma$ shows a negative correlation while $\tau$ shows the opposite dependency.
This competing correlation is why we see nonmonotonic profiles in $k_{nn}(k)$ and $c(k)$.
For some range of $k$, the negative correlation of $\sigma$ plays a major role while the positive correlation in $\tau$ becomes more evident in other region of $k$.
Figure~\ref{fig:compare_sim_eq_vec}(d) also shows the theoretical values, which completely agrees with the simulation results.

\section{Correlated hidden variables}\label{sec:correlated_h}

So far we have derived the analytic forms for various network properties under the assumption that $h$s are independent of each other in the original network. Here, we consider a more realistic case where the neighboring $h$s are correlated.
One of the typical examples of the correlated attributes is the homophily mechanism in social networks, meaning that people tend to form ties between those similar to themselves~\cite{mcpherson2001birds}.

When $h$s are correlated, one can conduct a rigorous calculation under a limited condition that $h$ is Markovian, i.e., the probability distribution of $h$ is conditional only on their neighbors' hidden variables.
We also assume that $h$ is independent of the local network topology, such that the degree, the average neighbor degree, and the local clustering coefficient are independent of $h$ of the node.
Under this assumption, it is straightforward to formally write down the equations for $P(k)$, $k_{nn}(k)$, and $c(k)$ as shown shortly.
Although these assumptions limit the applicability of the theory as the hidden variables are usually not Markovian, it serves as a good approximation for various practical cases and gives an idea about how the correlation in $h$ would affect the topology of the sampled networks.

When $h$ is correlated, the sampling probability around the node $\bar{r}(h)$ is written as
\begin{equation}\label{eq:r_bar_h_corr}
    \bar{r}(h) = \int dh' p_o(h'|h) r(h,h').
\end{equation}
Using Eq.~(\ref{eq:r_bar_h_corr}) instead of Eq.~(\ref{eq:r_bar_h}), the other equations in Sec.~\ref{subsec:degree} are still valid.
The degree distribution is calculated using Eq.~(\ref{eq:P_k}).

To calculate $k_{nn}(k)$, we have to replace Eq.~(\ref{eq:p_hprime_given_h}) by the following formula:
\begin{eqnarray}\label{eq:p_h_given_h_corr}
p(h'|h) = \frac{r(h',h)p_o(h'|h)}{\bar{r}(h)},
\end{eqnarray}
with the same $\bar{r}(h)$ as in Eq.~(\ref{eq:r_bar_h_corr}).
The remaining calculations in Secs.~\ref{subsec:knn} and \ref{subsec:ck} are the same.
Therefore, the degree correlation in a sampled network is written as the joint effect of the original degree correlation $p_o(\kappa'|\kappa)$, the correlation of $h$ in the original network $p_o(h'|h)$, and the sampling-induced assortativity.

To demonstrate how the correlation in $h$ works, we study the following model as a case study. The original network is constructed using the stochastic block model (SBM) where $N$ nodes are equally partitioned into $C$ communities of size $N_{C} = N/C$. The probability of making intracommunity and intercommunity links are given by $p_{\rm in}$ and $p_{\rm out}$, respectively, which are independent of the community. The hidden variable $h$ of a node in community $I$, where $I$ is the index of the community ranging from $1$ to $C$, is drawn from $\rho_{I}(h)$.
Thus, the distribution of $h$ for all the nodes is $\rho(h) = \sum_{I} \rho_{I}(h) / C$.
For $\rho_{I}(h)$, the same functional form as Eq.~(\ref{eq:weibull_rho_h}) with $\alpha=1$ is adopted, but with $h_0$ dependent on the community as $h_0(I) = 0.01I$. With this community dependent $h_0$, the nodes in the same community have similar $h$ compared to the nodes in other communities, yielding a positive correlation between neighboring $h$s.
For the sampling probability, Eq.~(\ref{eq:genmean_r}) with $\beta=-\infty$, i.e., $r(h_i,h_j) = \min\{h_i,h_j\}$, is used.

For this model, the conditional probability $p_o(h'|h)$ is given as
\begin{eqnarray}
p_o(h'|h) &=& \frac{ p_o(h',h) }{ \rho(h) },
\end{eqnarray}
where $p_o(h',h)$ denotes the probability that a link in the original network connects nodes of $h$ and $h'$.
This joint probability is given by
\begin{eqnarray}
p_o(h',h) &=& \sum_{I}\frac{\rho_{I}(h)}{C} \left[ q_{\rm in}\rho_{I}(h') + \frac{q_{\rm out}}{C-1}\sum_{I'\neq I}\rho_{I'}(h') \right],\nonumber\\
\end{eqnarray}
where $q_{\rm in}$ and $q_{\rm out}$ are the fractions of the intra- and intercommunity links, respectively.
These are given by
\begin{eqnarray}
q_{\rm in} &=& \frac{p_{\rm in}(N_{C}-1)}{p_{\rm in}(N_{C}-1)+p_{\rm out}(N-N_{C})},  \\
q_{\rm out} &=& \frac{p_{\rm out}(N-N_{C})}{p_{\rm in}(N_{C}-1)+p_{\rm out}(N-N_{C})}.
\end{eqnarray}


\begin{figure}[htbp]
  \begin{center}
    \includegraphics[width=\columnwidth]{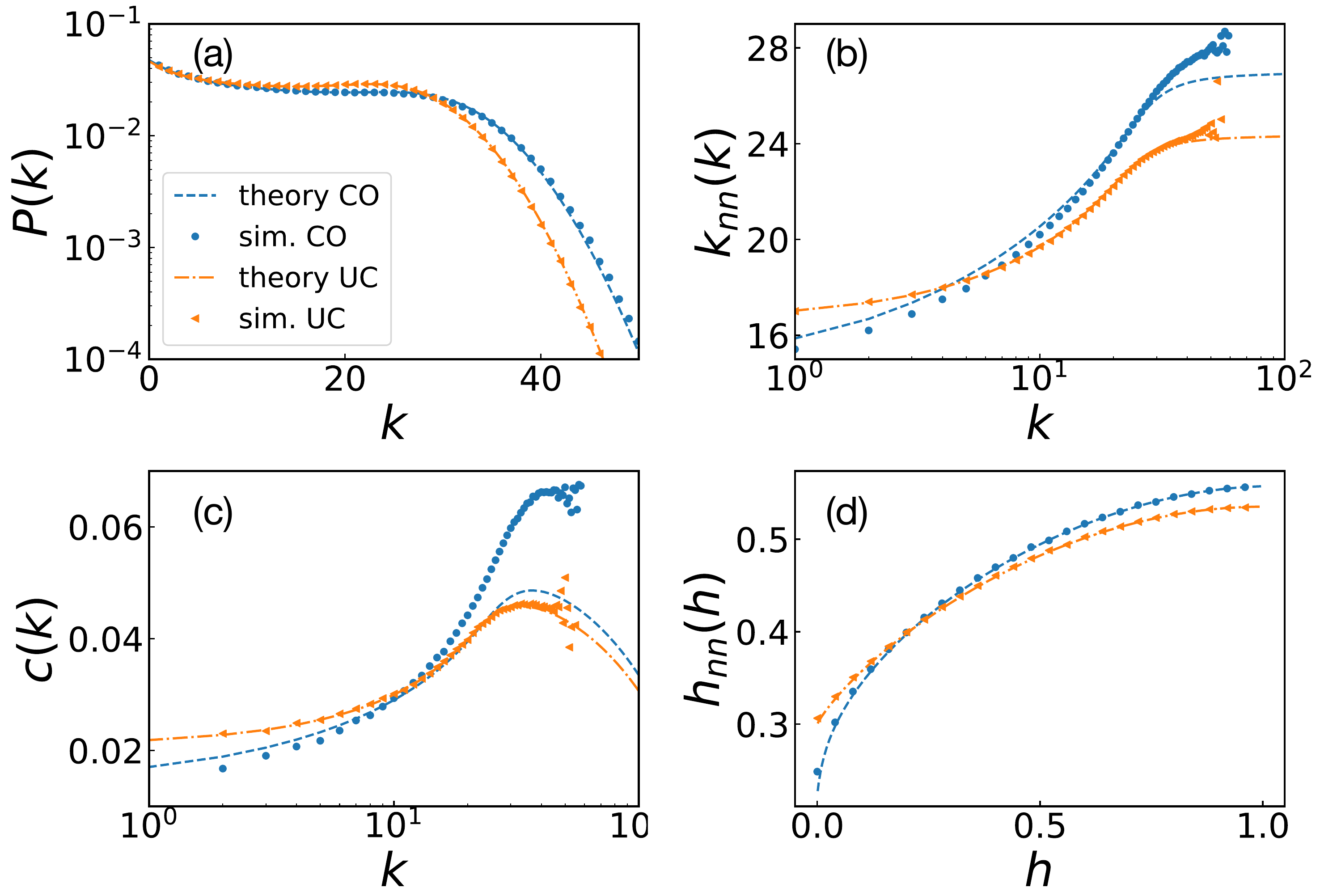}
    \caption{
    (a) Degree distribution $P(k)$, (b) average neighbor degrees $k_{nn}(k)$, (c) clustering spectrum $c(k)$, (d) correlations of neighboring hidden variables in the sampled network for the model with correlated $h$.
    Simulation results, depicted by symbols, are compared with the theoretically predictions, depicted as dashed curves.
    The original network is constructed using the stochastic block model where $N=10000$, $C=100$, $p_{\rm in} = 0.5$ and $p_{\rm out}=50/9900$.
    For each community, $h$ is drawn randomly from a distribution whose mean is dependent on the index of the community to introduce the correlation.
    The results are compared with a null model where the correlation is removed by shuffling $h$.
    The correlated case and the uncorrelated cases are denoted as CO and UC in the legend, and drawn in blue and orange, respectively.
    The fraction of the sampled links are $16\%$ and $15\%$ for the correlated and uncorrelated cases, respectively.
    The simulations results are averaged over $200$ independent runs.
    }
    \label{fig:corr_compare_sim_eq}
  \end{center}
\end{figure}

Figure~\ref{fig:corr_compare_sim_eq} shows the simulation results as well as the theoretical predictions for this model. The parameter values of $N = 10000$, $C=100$, $N_C = 100$, $p_{\rm in}=0.5$, and $p_{\rm out}=50/9900$ are used.
With this setting, half of the links are intra-community links while the other half of the links are made between different communities.
To investigate the effect of the correlations, the uncorrelated version of the model was also studied.
For the uncorrelated model, $h$ of the nodes are randomly shuffled while the other settings kept the same.

Figure~\ref{fig:corr_compare_sim_eq}(a) shows the degree distribution for both correlated and uncorrelated cases. When the correlation is introduced, the degree distribution has a heavier tail than that for the uncorrelated case. This is because a node with a high $h$ tends to be surrounded by nodes with higher $h$.
High-degree nodes have even higher degrees when $h$ is correlated.
Figure~\ref{fig:corr_compare_sim_eq}(d) shows the correlation of $h$ in the sampled networks for the correlated and uncorrelated cases.
The correlation in the original network enhances the positive correlation in the sampled network. The effects of the correlations are also observed in $k_{nn}(k)$ and $c(k)$ for both correlated and uncorrelated cases.
The degree assortativity and the increasing behavior of $c(k)$ get stronger for the correlated case as shown in Figs.~\ref{fig:corr_compare_sim_eq}(b) and \ref{fig:corr_compare_sim_eq}(c).

The theoretical predictions using Eqs.~(\ref{eq:r_bar_h_corr}) and (\ref{eq:p_h_given_h_corr}) are also shown in Fig.~\ref{fig:corr_compare_sim_eq}.
The theoretical curves for $P(k)$ and $h_{nn}(h)$, shown in Figs.~\ref{fig:corr_compare_sim_eq}(a) and \ref{fig:corr_compare_sim_eq}(d), agree very well with the simulation results.
To calculate the sampled degree around a node, $h$s of the focal node and its surrounding nodes are necessary.
Because longer correlations are not necessary, these equations, that take the neighboring correlations into account, are rigorous.
However, the theoretical curves for $k_{nn}(k)$ and $c(k)$ show deviations from the simulation results as depicted in Figs.~\ref{fig:corr_compare_sim_eq}(b) and \ref{fig:corr_compare_sim_eq}(c).
This is because $k_{nn}(k)$ and $c(k)$ depend on longer correlations. The average neighbor degree depends on $h$s of the neighbor node and its neighbors, i.e., the correlations between the focal node and its next nearest neighbors affect $k_{nn}(k)$. The clustering coefficient also depends on the correlations between the next nearest neighbors.
A theory which takes into account the correlations of $h$ between next nearest neighbors would improve the accuracy.
Even though these equations are not rigorous ones, they give good approximations practically and tell how the correlations affect the sampling.

\section{Summary and Discussion}\label{sec:conclusion}

In this paper, we have studied a class of sampling on networks, where a sampling probability of a link depends on the attributes of the connected nodes.
This is typically the case when a multiplex is sampled by some layers only.
The rigorous results for $P(k)$, $k_{nn}(k)$, and $c(k)$ for the sampled network are shown for general functional forms of $\rho(h)$ and $r(h_i,h_j)$ when the attributes of the nodes are independent.
The analytic calculations are found to be in good agreement with the Monte Carlo simulations.
As shown in Table~\ref{tab:eq_summary}, the properties of the sampled networks are written as the aggregate of the contributions from the original network and the hidden variables.
The theory also presents how quantities in the sampled network depend on the original network.
For instance, it is shown that the sampled degree distribution $P(k)$ depends on the original degree distribution $P_o(\kappa)$ but not on the higher order correlations.

As a concrete example, we have studied the model where $r(h_i,h_j)$ is given as the generalized mean of $h_i$ and $h_j$, which was proposed to model sampling a communication channel of the social network~\cite{torok2016big}.
Using the equations in Table~\ref{tab:eq_summary}, we compare the sampled networks and the original networks in various aspects. For this model, the original network property is manifested only for the limited range of $k$, indicating that recovering the original network from the sampled network is unfeasible.
One of the lessons learned from this example is that the network we observe does not necessary reflect the original network properties. Instead, it may reflect the attributes of nodes.

We also present a theory for the case where the neighboring $h$s are correlated as in reality the attributes are not independent.
Although the theory is not rigorous for general cases, it gives a good approximation in practical cases and tells how the correlation in $h$ alters the properties of the sampled networks.

So far, we have limited ourselves to the case where the original network properties and the hidden variable are uncorrelated, by assuming that $h$ and the network quantities ($\kappa$, $\kappa_{nn}$, and $c_o$) are uncorrelated.
Although we leave the study for the correlated case for future researches, its impact could be highly significant.

Another future research issue would be the method development to infer the original network and/or $h$ from empirical data sets. As in Refs.~\cite{newman2016structure,hric2016network}, the usage of metadata, working as a proxy of $h$, would be of great help because the correlation of $h$ in the sampled network is independent of the topology of the original network.
It will be helpful to infer the functional form of $\rho(h)$ and $r(h_i,h_j)$ as well.
We believe our results serve as the basis for these future researches.

\begin{acknowledgements}
Y.M. acknowledges support from MEXT as ``Exploratory Challenges on Post-K computer (Studies of multi-level spatiotemporal simulation of socioeconomic phenomena)'' and from Japan Society for the Promotion of Science (JSPS) (JSPS KAKENHI; Grant no. 18H03621).
H.-H.J. was supported by Basic Science Research Program through the National Research Foundation of Korea (NRF) funded by the Ministry of Education (NRF-2018R1D1A1A09081919).
Y.M., H.-H.J., J.T., and J.K. are thankful for the hospitality of Aalto University.
The source code for the analysis is available at~\url{https://github.com/yohm/sim_sampling_analysis}.
\end{acknowledgements}

\bibliography{sampling}

\end{document}